\documentclass[twocolumn]{article}

\usepackage[utf8]{inputenc}		%
\usepackage[T1]{fontenc}		%
\usepackage{graphicx}			%
\usepackage{amssymb} 			% Symbols
\usepackage{amsmath} 			% Symbols
\usepackage{wasysym}
\usepackage{url}
\usepackage[sort&compress,numbers]{natbib}
\usepackage{fullpage}
\begin{document}

\title{Could the Well of an Orbital Lift be used to Dump Greenhouse Gases into Space?}

\author{O. Bertolami\footnote{Email address: orfeu.bertolami@fc.up.pt}}

%\address{Departamento de Física e Astronomia and Centro de Física das Universidades do Minho e do Porto,\\ Faculdade de Ciências, Universidade do Porto,\\ Rua do Campo Alegre s/n, 4169-007 Porto, Portugal}

\date{\small{Departamento de Física e Astronomia and Centro de Física das Universidades do Minho e do Porto,\\ Faculdade de Ciências, Universidade do Porto,\\ Rua do Campo Alegre s/n, 4169-007 Porto, Portugal}}

\maketitle

\begin{abstract}
\noindent
Anthropogenic greenhouse gases have been changing significantly the climate and causing dire effects on the dynamics of the Earth System. We examine the conditions under which the well of a geostationary orbital lift can be used to dump greenhouse gases into space. 
\end{abstract}

%-----------------------------------------------------------------------------%
%-- Body ---------------------------------------------------------------------%
%-----------------------------------------------------------------------------%

\section{Introduction}
\label{sec:introduction}

Anthropogenic climate change due to the accumulation of greenshouse gases in the atmoshere is the most serious civilizational threat of our time. Its solution demands for a radical change on the tennets of the consumption societiy driven by cheap fossil fuels and built upon the mistaken assumptions that Earth's resources are limitless and that the planet is an inifinte dump of waste. Obviously, any lasting fix of the climate change involves a dramtic reduction of the emissions of greenhouse gases and profound socio-economic changes. Nevertheless, it is important to realise that the problem must be addressed in little more than a decade or so and that we may be running out of time to carry out encompassing long term changes. In this context, adaptation and mitigation strategies are in the class of the absolutely minimal set of necessary measures in order to get us some extra time to fix the problem. These include, besides the urgent measures to act in situations of climate emergency caused by droughts, heat waves, flooding, wild fires, etc, the acceleration of efforts to decarbonise human activities, extending and generalising the use of renewable energies, setting up means to carbon capture by afforestation, restoration of ecosystems and other chemical-mechanical means, besides rational use of water, vital resources, etc. 

Besides the abovementioned conventional measures, some more controvertial ones such as ocean fertilisation and alkalinity enhancement have also been considered. Other geoengineering proposals include, for instance, albedo enhancement through passive daytime radiative cooling \cite{Zeven1,Wang}, the use of sky-facing thermally-emissive surfaces to radiate heat back into space \cite{Chen,Munday}, stratospheric aerosol injection (SAI), the so-called ``Budyko blancket"  \cite{Budyko,Crutzen,Rash,Lenton}, cloud brightening or a large set of mirrors in the sky to reflect back into space a fracion of the incoming solar irradiation (see Ref. \cite{Lenton} for a review). Relevant steps towards a better understanding the way aerosols grow at high altitude through the CLOUD experiment at CERN \cite{Wang2} and $CO_2$ conversion via coupled plasma-electrolysis \cite{Goede,Pandiyan} might turn out to be interesting avenues for mitigation strategies in the future. Of course, any geoengineering proposal involves some amount of negative side effects. 

Some proposals of geoengineering consider change in the illumination conditions of the Earth by the Sun through space reflectors. A space mirror \cite{Early,Roy} and a myriad of reflecting bubbles \cite{MIT} have been proposed, but these are somewhat radical forms of intervention as they affect the whole electromagnetic spectrum of the incoming solar irradiation. These devices are supposed to be located at the L1 point in order to be unaffected by the gravitational forces of the Earth and Sun. 

In fact, any device that traps and reflects predominantly thermal radiation might be useful to reduce 
the amount of infrared radiation traping in the atmosphere. Hence, an hypothetical device could involve a transparent vessel filled with infrared traping gases or materials in a suitable orbit in order to deplet the incoming infrared radiation while allowing that radiation with the remaining wavelegths could travel through. The infrared shadow of this device, for instance in a geostationary orbit, could allow for the reduction of the infrared
radiation in a specific region \cite{OB}. A figure of merit of $1.6\%$ of overall reduction is often referred to in order to have an impact on the continuous climbing of the global temperature. This proposal will be discussed elsewhere.

In this brief note, we examine the feasibility of using the well of a geostationary orbital lift, more popularly known as space elevator, for dumping greenhouse gases into space. We shall assume that the known constraints on the setup of a orbital lift are met and focus on the efficiency of using this infrastructure as a device to dump greenhouse gases away from Earth's atmosphere.

An orbital lift, space elevator or space bridge was conceived long ago by Konstantin Tsiolkovsky in 1895 and is often depicted in science fiction as a method to reach space. It consists of a tether anchored to Earth's surface close to the Equator and a counterweight that extends itself into space beyond a geostationary orbit ($r_G= 35786$km). This configuration allows that the gravity force and upward centrifugal force balance each other. Of course, the feasibility of the concept depends crucially on the capability of the mateirals involved in the structure to hold the required stress and having the compressive strength to support its own weight. 

In 1959, the Russian engineer Yuri Artsutanov proposed that it would be more realistic to use a geostationary satellite as the base from which one could deploy the structure of the orbital lift downward \cite{Artsutanov}. In 1960's and 1970's American engineers have discussed similar concepts and reached the conclusion that the needed strength of the structure's materials would have to be at least two times thougher than the ones hitherto known: graphite, quartz and diamond \cite{Isaacs}. It was also pointed out that a cross-section-area profile that tapered with the altitude would be more suitable for an orbital structure \cite{Pearson}.

More recently, several iniciatives and competitions have appeared aiming to revitalise the orbital lift concept stimulated by advances in material science, more particularly, in knowledge acquired in the development of carbon nanotubes. 

Indeed, in 2019 the International Academy of Astronautics published a report \cite{IAA} assessing the state of art on the matters related to the orbital lift, stressing that it might be a reality in the near future given developments on the manufacturing of macro-scale single crystal graphene, whose specific strength is actually higher than the one of carbon nanotubes.

Thus, given that it is believed that setting up an orbital lift structure is not completely impossible, it is not at all futile to consider the possíbility of using this space device to dump into space the excess of greenhouse gases due to anthropogenic activities. In what follows we shall consider the $CO_2$ case.

\section{The Basic Features of the Proposed System}
\label{sec:model}

As stated above, we shall assume that the structure of the orbital lift is within reach and consider the well of its structure. which extends upwards up to $r_G \simeq 35786$ km, Let us consider that it has, for simplicity, a constant cross-sectional area, $A=\pi r^2$, where $r$ is the radiius of the well. The anchor of the orbital lift can be a geostationary bulky satellite in an equatorial plane orbit and whose struture can be built downwards as suggested by Artsutanov. 

Once the body of the lift is constructed, the idea is to inject $CO_2$ into the well of the orbital lift and create an upward flow that allows for dumping $CO_2$ into space. Of course, natural conditions do not allow for any effective upwards flow as Earth's escape velocity is much higher than the typical average velocities of the molecules that compose the air. Furthermore, atmosphere's density decays exponentially and its temperature profile as a function of the altitude is complex\footnote{The temperature profile of the atmosphere along its layers is approximately as follows: at the troposhere (0 to 10 km) the temperature drops linearly from the surface temperature to about $-55^{0}C$; it remains constant at about $-55^{0}C$ at the tropopause (10 to 20 Km); it increases linearly at the stratospehre (20 to 45 km) till about $0^{0}C$;  it decreases linearly at the mesosphere (60 to 90 km) till $-80^{0}C$; it raises again at the ionosphere (60 to 700 km), most particularly beyond the von K\'arman line (100 km).}. Thus, conditions for an upward flow must be created and hence the well of the lift must be sealed and its conditions cannot be the atmospheric ones. This means that the first steps of the operation are to pump out the air of the well and inject $CO_2$ in its interior. Transporting the $CO_2$ upwards can be achieved through its ionisation and an applied electric field with the right polarity. This will create an upwards dynamical flow. Hence, the necessary conditions to setup an upward flow can be realistically achieved through the following steps: i) pumping out the air inside the well of the orbital lift;  ii) separation of the $CO_2$ in the air; iii) injection into the well of the accumulated $CO_2$; iv) ionisation of the $CO_2$ in the well; v) acceleration of the charged $CO_2$ through an electric field along the vertical axis of the orbital lift. The broad technical features of these steps are described below. It is relevant to point out that we aim, in its original version, to keep our device as simple as possible. 

\vspace{0.1cm}
\noindent
i) The air in the well is pumped out till it reaches a density $10^{-4}$ smaller than the atmospheric one;    

\vspace{0.1cm}
\noindent
ii)  It is known that $CO_2$ diffuses in porous media (see Ref. \cite{diffusion} for a review). Thus, it is quite feasible to built up a high concentration of $CO_2$ with a somewhat uniform distribution along the low altitude section of the orbital lift through the diffusion processes that separate the $CO_2$ in the air. This can be carried out intensively at the bottom of the lift using various sources of $CO_2$ or throughout a series of diffusive processes along the low altitude part of the troposphere, the denser part of the atmosphere;

\vspace{0.1cm}
\noindent
iii) Injection can take place ithrough mechanical pumping or via a pressure gradient between the separation reservoir and the well. The required density of $CO_2$ is about $4 \times 10^{-4}kg/m^3 $. This procedure is straightforward and, in principle, does not require any major innovation or technological breakthrrough. The extrenal surface of the orbital lift is quite large, $A_{OL} \simeq 2 \pi r r_G$, and can be used to absorb solar radiation which can be photovoltaically converted into electric energy and heat the gas. The $CO_2$ freezing point is $T=194.65~K$, so the temperature inside the well must be kept above the freezing point. As will be seen below, the $CO_2$ can be mixed with some other gas;  

\vspace{0.1cm}
\noindent
iv) Ionisation of the $CO_2$ as a method of separtion was proven feasible long ago \cite{Ito}. This means that ionisation can also be used in the processes i) and ii) described above. However, in order to convey our concept in the simplest possible way, we shall keep the steps enumerated above separate from each other. This means that in principle there is plenty of room to optimise our concept. In Ref. \cite{Ito} seperation of $CO_2$ from a mixture with an inert gas ($He$) was shown to be effective. Ionisation was achieved through irradiation by soft X-ray. It was reported that some $CO_2$ was decomposed, but it was found that separation with a maximum efficiency was obtained up to certain concentration of $He$ ($14 \%$) for an applied voltage of $600~V$. We retain from the study reported in Ref. \cite{Ito} that $CO_2$ can be ionised and hence can be accelerated by an electric field. The reported results indicate that for a concentration of $5 \times 10^{19}$ molecules/$m^3$ of $CO_2$, the amount of ionised molecules was six orders of magnitude smaller, meaning that the charge to mass ratio is typically about $q/M = 1.37 \times 10^{19} |e| \simeq 2.2C/kg$, where $e= -1.6 \times 10^{-19}~C$ is the electron charge;

\vspace{0.1cm}
\noindent
v) Thus, once an amount of $CO_2$ is in the well of the orbital lift (processes i) and ii)), it can be ionised and accelerated upwards through an electric field. Assuming that an aggregate of charged $CO_2$ has a vanishing initial velocity in the vertical direction, once the electric field is applied, after a height, $H$, it will have a final velocity, $v_f$:
\begin{equation}
v_f = \sqrt{2(\eta -1)gH},
    \label{eq:vf}
\end{equation}
where $g$ is the acceleration of gravity, $\eta= qE /Mg$, $E$ being the applied electric field and $M$ the mass of the ionised aggregate of $CO_2$ molecules. Clearly, $v_f$ must be at least as large as Earth's escape velocity, $v_E \simeq 11.2~km/s$. 

Once the velocity reaches the value Eq. (\ref{eq:vf}), the aggreagate will climb a distance $x=(\eta-1) H$ in a region where the electric field vanishes, till its velocity drops to zero. The aggregate can be then be submitted to an electric field again as described above. Excluding the single section configuration, where the electric field extents over the whole structure of the orbital lift, which might be too demanding technically, the workable configurations involve: the first and last sections of the welll under the effect of the electric field and a middle section with no electric field (Scenario 1); or three sections with an electric field and 2 intermediate sections with no electric fields (Scenario 2). Other configurations, for instance, with 4 sections with an electric field and 3 sections with no electric field do not allow for the ionised lump of $CO_2$ to reach the escape velocity.

For a voltage per metter of about, say $10~V/m$, just slightly higher than typical values used in long transmission lines of electricity, then $\eta \simeq 2.2$ and $H_1 \simeq 11.2 \times 10^6~m$, $x_1 \simeq 13.4 \times 10^6~m$ and $v_{f1} \simeq 15 ~km/s$ for Scenario 1. For the Scenario 2, one gets: $H_2 \simeq 6.6 \times 10^6~m$; $x_2 \simeq 7.9 \times 10^6~m$; and $v_{f2} \simeq 11.9 ~km/s$ 

The outward flow of $CO_2$ can be estimated as $\Phi = j \pi r^2$, where $j= \rho v_f$. For $\rho= 4 \times 10^{-4} ~kg/m^3$ and $r = 15 ~m$ one gets for Scenario 1, $\Phi_1 = 4.2~ ton/s$. This means that over a year, about $1.31 \times 10^8$ tons can be dumped into space. This is about $2 \%$ of the anthropogenic $C0_2$ generated over the same period ($6.4 \times 10^9~ton/year$)\footnote{If the electric field extended over the whole structure of the orbital lift, the resulting outward annual flux would be about $3.6 \%$ of the antropogenic generated $CO_22$.}. For Scenario 2, one gets:  $\Phi_2 = 3.4~ ton/s$ or $1.04 \times 10^8$ ton/year ($1.6 \%$ of the antrhropogenic amount). These are relativily modest amounts, but indicate that if an orbital lift is built its well can be used, under the conditions discussed above, as a device to dump $CO_2$ into space. Notice that under standard conditions of temperature ($T= 273.15~K$) and pressure ($p=1.013 \times 10^5~Pa$), the density of the $CO_2$ is $\rho_{STP}= 1.96~kg/m^3$, so the chosen density of $CO_2$ is a factor $5 \times 10^3$ smaller. Assuming that the flow is incompressible, the dynamical upward pressure in the sections with an electric field is about $p_1= 4.5 \times 10^4 ~Pa \simeq 0.44~atm$ for Scenario 1 and $p_2= 2.8 \times 10^4 ~Pa \simeq 0.28~atm$ for Scenario 2. Of course, improvements on the ionisation rate would allow for much better performances of the concept for a lesser dense amount of $CO_2$ at the first section of the well. For the ionisation rate of Ref.  \cite{Ito}, it is required that the initial density of $CO_2$ is about 500 times greater than the normal conditions. 
 
Naturally, in principle, similar manipulations can also be used for handling methane, a potent greenhouse gas whose concentration in the atmosphere has been sharply increasing due to the farming indutrtry and the hydraulic fracturing (fracking) technique for extracting gas and oil from shale rock.

Before closing this section, let us discuss two direct physical implications of the proposed set of operations of our device. If properly handled, these effects do not affect the performance of our device, but, for sure, they deserve being discussed\footnote{I am thankful for Clovis de Matos for pointing them out.}. The first one concerns the Lorentz force due to the upward flow of charge. The generated magnetic field, $B$, can be estimated by Amper\'e's law: $B=\mu j r/2$, where $\mu$ is the magnetic permeability constant of the $CO_2$ gas, which given its low density we shall assume to be close to the vacuum value, that is: $\mu=\mu_0= 4\pi \times 10^{-7}N/A^2$. Considering the most demanding scenario (scenario 1), the resulting Lorentz force, $|\vec{F_L}|=j \pi r^2 l B$, where $l$ is a length scale, which for a negatively charged gas is outward and about $3.16 \times 10^3~N$. This yields a negligible outward pressure for the electrified sections of the well ($l=H$): $3 \times 10^{-6}~Pa$. 
As for the effect of Earth's magnetic field, whose strength is about $(25-65) \times 10^{-6}~T$, assuming it has only a north direction component, the corresponding Lorentz force is inward and about the same order of magnitude of the effect generated by the flow of $CO_2$. Hence, at the section of the well with an electric effect, the total Lorentz force approximately cancells out, while it is about as small as the Lorentz force computed above, except that it is predominatly inward, at the sections of the well with no electric field.  

The second effect is the thrust due to the injection of gas into space that is transmitted on the structure of the space lift. The dominant term is given by $\Phi v$ which is, for scenario 1, about $6.3 \times 10^7~N$. This can impose a considerable extra strain on the structure of the orbital lift. In order to avoid this undesirable effect, a simple solution is to consider a symmetric ejection of the $CO_2$ along a direction perpendicular to the axis of the well. This cancellation can be achieved through a radially symmetric set of nozzles perpendicular to the axis at the top end of the well that delivers the gas away from the structure of the lift.  Actually, the ejected $CO_2$ could be used to fill the infrared absorbing vessels mentioned above and whose details will be presented somewhere else \cite{OB}.

%-----------------------------------------------------------------------------%

\section{Discussion and Outlook}

Uncontroversial evidence indicates that a climate crisis is unfolding. Its cause is anthropogenitc and it puts the habitability of the planet under threat. In fact, the rise of the global temperature due to the continuous climbing of the concentration of greenhouse gases are driving the Earth System (ES) to a Hot House Earth State were all the major regulatory ecosystems can reach their tipping points \cite{Steffen2018}. Moreover, theoretical predictions based on a physical model and on the ensued Anthropocene equation show that a Hot House Earth State is an inevitable outcome given the present intensity of human activitites (see e.g. Refs.  \cite{OB-FF1,OB-FF2,OB-FF3,OB-FF4,AB-OB}).

Indeed, the methodology proposed in the above references to describe the ES can be used as a classification scheme for rocky planets  \cite{OB-FF5} and the resulting analysis shows that Venus is in fact in a Hot House Earth like state. In other words, Venus is very much like an Earth with an uncontrolled $CO_2$ problem. This resemblance stresses the likelihood that the Anthropocene is a transition between the Holocene to a much hotter Venus-like Earth. The dynamical system analysis of the Anthropocene equation emerging from the model of Ref. \cite{OB-FF1} confirms that this hotter Venus-like state is indeed an attractor of trajectories \cite{OB-FF2} and may be driven, under conditions, into a chaotic regime \cite{AB-OB}. This emphasises the importance of setting up strategies to mitigate the effect of the excess of greenhouse gases in the atmosphere.  

The design of various geoengineering projects have been proposed to mitigate the ongoing climate change crisis. In this work we have suggested that the well of an orbital lift, a structure that has been primarily proposed to reach space, can be used as a geoengeering device to dump modest amounts of $CO_2$ into space. We argued that many of the tecnological steps towards achieving this goal have already been mastered, but the hurdle of constructing the orbital lift itself. The latter seems to be still in the realm of science fiction. In any case, we believe that it is relevant to point out that an extraordinary device such as the orbital lift can also be used as a tool to face the most troubling civilisational challenge of our time. We have shown that through quite feasible steps, the well of the orbital lift can be used to dump modest amounts of $CO_2$ into space. We have discussed the requirements to transport $CO_2$ till space and estimated the flux of $CO_2$ that can be dumped into space. For sure, keeping a constant density of $CO_2$ along the well and a substantioal fraction of it ($10^{-6}$) uniformly ionised is somewhat challenging, but not at all impossible. 

Of course, it is well understood that any proposal to remove $CO_2$ from the atmosphere, and our proposal is no exception, is dwarfed by the pantagruelic antropogenic emission. This means that most of the resources to combat climate change must be geared towards a significant decarbonisation of the human activities. On its hand, this implies that a drastic reduction of the consumption patterns of our society must take place. A coupled effort must also be made in changing the brutal and disfunctional way the existing market economy destroys ecosystems. The long term habitability of the planet for all species is under threat. It is already quite clear that the only realistic way towards a sustainable future is through a rational and insightful economic degrowth.

%\vspace{0.5cm}

%{\bf Acknowledgments~~}

%\noindent
%This work is partially supported by Funda\c{c}\~ao para a Ci\^encia e a
%Tecnologia (Portugal) under the project POCI/FIS/56093/2004.

%\vspace{0.3cm}

%\vfill
%\newpage

\bibliographystyle{unsrtnat}

\end{document}